\documentclass[aps,pre,twocolumn,showpacs,superscriptaddress,groupedaddress]{revtex4}
\usepackage{graphicx}  % needed for figures                                                           
\usepackage{dcolumn}   % needed for some tables                                                       
\usepackage{bm}        % for math                                                                     
\usepackage{amssymb}  
\begin{document}

\title{Transport coherence in a time-asymmetric rocked ratchet model}
\author{Mamata Sahoo}
\email{mamata.sahoo@iisertvm.ac.in}
\affiliation{School of Physics, Indian institute of science education and research, Thiruvananthapuram -695016, India}
\author{A. M Jayannavar}
%\email{jayan@iopb.res.in}
\affiliation{Institute of Physics, Sachivalaya Marg, Bhubaneswar-751005, Orissa, India}
\affiliation{Homi Bhaba National Institute, Training School Complex, Anushakti Nagar, Mumbai 400085, India}

%\author{Mamata Sahoo$^{1'*}$, A. M Jayannavar$^{2'3}$}
%\address{
%$^1$ School of Physics, Indian institute of science education $\&$ research, Thiruvananthapuram -695016, India\\
%$^2$ Institute of Physics, Sachivalaya Marg, Bhubaneswar-751005, Orissa, India\\
%$^3$ Homi Bhaba National Institute, Training School Complex, Anushakti Nagar, Mumbai 400085, India
%}}

%\ead{$^*$mamata.sahoo@iisertvm.ac.in, jayan@iopb.res.in}
\begin{abstract}                          
We study the dynamics of an over damped Brownian particle in a saw tooth potential in the presence of a temporal asymmetric driving force. We observe that in the deterministic limit, the transport coherence, which is determined by a dimensionless quantity called Peclet number, $Pe$ is quite high for larger spatial asymmetry in the ratchet potential. For all the regime of parameter space of this model, $Pe$ follows the nature of current like Stokes efficiency. Diffusion as a function of amplitude of drive shows a minimum exactly at which the current shows a maximum. Unlike the previously studied models, the $Pe$ as a function of temperature shows a peaking behavior and the coherence in  transport decreases for high temperatures. In the nonadiabatic regime, the $Pe$  as a function of amplitude of drive decreases and the peak gets broader as a result the transport becomes unreliable.
\vskip.5cm
%{\small{\textit{PACS :} 05.40.-a; 05.70.lw}} 
%{\small{\textit{Keywords:}Ratchets, entropy production, noise, energy loss.}}
\end{abstract}
%\end{frontmatter}


%\noindent{\it keywords}:Ratchets, Efficiency, Fluctuations

\maketitle

\section{Introduction}
In the past years, there is an increasing interest in the study of non-equilibrium thermodynamics due to its relevance in the field of nano-technology and biology \cite {Bustamante, Ritorts}. One of the great challenges in the field of nano-technology is the design and construction of microscopic version of motor/devices that can perform directed motion using input energy and do some mechanical work. Driving such directed motion is what protein motors (a kind of Brownian motor/molecular motor) do over the course of years by evolution in every cell in living bodies \cite{Blocks}. It has been realized that such unidirectional motion inside cell is possible only due to the presence of noise. Noise is everywhere in nature and it plays a creative and stabilizing role in the dynamics of many nonequilibrium phenomena, e.g., noise induced directed transport, stochastic resonance, noise induced phase transition, noise induced stability of unstable states, noise indued ordering \cite{Hangii, Julicher1, Astumian, Astumian1, reimann, Reimann,jayan, Gammaitoni} etc. In the recent years, the phenomenon of "noise induced directed transport" is being studied extensively both in nano-physics and biology. The operating principle of Brownian motors/Brownian ratchets is based on the principle of noise induced directed transport \cite{Reimann}. Various kinds of Brownian ratchet models like flashing ratchets, rocking ratchets, time-asymmetric ratchets, and inhomogeneous ratchets have been discussed based on the various ways of introducing the asymmetry in the ratchet system \cite{Reimann}. The theoretical concept of noise induced transport in case of Brownian motors/Brownian ratchet models have been realized experimentally in a variety of systems. The examples include cold atoms in optical lattices \cite{Jones}, colloidal particles in holographic optical trapping patterns, \cite{Lee} ratchet cellular automata \cite{Babie} etc.

The most basic measure for characterizing the motion of transport in the case of a Brownian motor or Brownian ratchet model is it's average velocity in the long-time limit. This describes how much time a motor requires to overcome a given distance in the asymptotic regime. However, velocity is not only the measure for characterizing the nature of the transport. The other attributes are the quality and the effectiveness of transport. This is because, the noise induced transport in a motor is always acompained by a dispersive spread or diffusion. If the diffusive spread is very large compared to the distance covered by it, the quality of transport degrades and the transport becomes unreliable. Similarly, in order to have an effective transport in a motor, one needs to look into the following questions: how much energy is input into the system?, out of which how much is going to be converted into directed motion?, and how much of it is spread out to the environment as dissipated heat?. A proper quantifier for characterizing this aspect of transport is the 'efficiency of energy transduction". Depending on the choice of the input and output energies in the performance of the motor, different definition of efficiency would characterize various aspects of energy conversion in its efficient operation. Since Brownian motor moves in a dissipative environment or highly viscous medium, a proper measure of efficiency in such viscously loaded motor is known as Stokes efficiency, which can be viewed as a measure of how efficiently a motor can utilize the input energy to drive a certain distance against the acting viscous force. Although, substantial progress have been done in the study of nature of currents and efficiency of transduction in various class of ratchet models \cite{Magnasco, Mahato, sekimoto, parrondo, Dan, Raishma, Makhnovskii, Derenyi, Munakata}, a few studies address the questions related to the quality or reliability of transport in ratchet models \cite{Freund, Linder, Machura, Schnitzner, Wang, krishnan1, Roy}. In some of our recent studies \cite{Raishma, Sahoo, Sahoo1}, it is observed that in a time asymmetric rocked ratchet model, the values of both the thermodynamic and Stokes efficiency are quite large in a suitable parameter regime as compared to the previously studied ratchet models and hence the time asymmetric rocked ratchet model can be treated as an efficient Brownian ratchet. However, the coherence or reliability of transport has not been reported so far in this particular ratchet model. In this study, we are mainly interested in the coherence or reliability of transport of an over damped Brownian particle in a temporally asymmetric rocked ratchet model. The transport of a Brownian particle in a ratchet model is said to be coherent or reliable, if the accompainig diffusion is much smaller than the distance covered by it. This, in turn, can be quantified by a dimensionless quantity, known as Peclet number, $Pe$ which is defined as the ratio of the average velocity to the diffusion coefficient. 

Originally, the concept of Peclet number arises in the problems of heat transfer in hydrodynamics where it stands for the ratio of heat advection to diffusion. When the Peclet number is small, the random motion dominates and when it is large the ordered and regular motion dominate. The value of the Peclet number depends on some characteristic length scale of the system. While dealing the ratchet motion, the most adequate choice for such a length scale is the period of the ratchet. For example, if the particle in the ratchet on an average covers a distance, $L$ due to its finite average velocity, $<v>$, the peclet number can be expressed as $Pe=\frac{<v>L}{D}$. In our model, we consider $L$ to be spatial periodicity of the ratchet potential. Higher Peclet number implies more coherence in the transport. In particular, quantitatively if the Peclet number is larger than $2$, the transport is said to be more coherent or reliable \cite{Freund}. The Peclet number for some of the flashing and rocking ratchet model shows very low values, i.e., $Pe \sim 0.2$ and $Pe \sim 0.6$, respectively resulting very low coherence in transport. However, in some of the studies it is found that a special kind of asymmetry helps in increasing the value of peclet number, i.e., Pe upto $20$ in a suitable regime of parameter space \cite{ Linder}. The experimental observation in Ref. \cite{Schnitzner} shows that $Pe$ in case of biological motors ranges from $2$ to $6$ and these motors perform very efficiently as well as coherently inside living cells. The coherence in transport was also observed in various driven model systems \cite{Roy}. In a very recent study, it is shown that the characteristic performance of a ratchet can arise from the complex interplay between different characteristic time scales of the system \cite{Viktor}. 

In the present work, we are mainly interested in a rocked ratchet model with a time asymmetric driving. Enhannced thermodynamic efficiency and Stokes efficiency were observed in this particular ratchet model in a suitable range of physical parameters \cite{Sahoo}. In the present study, we observe that in the deterministic limit, the larger spatial asymmetry in the potential enhances the coherence in transport, at the same time the temperature and driving frequency degrades the quality of transport. The structure of the paper is as follow. In Sec.II we introduce our model and the basic physical quantities of interest, namely, the average particle current, the effective difussion coefficient and the Peclet number. In Sec.III, we give the numerical simulation details of the model. The Sec.IV is devoted to the exploration of the transport properties in this particular rocked ratchet model. The results of our findings are briefly summarized in Sec.V.

\begin{figure}[hbp!]
\begin{center}
\includegraphics [width=3in,height=2.5in]{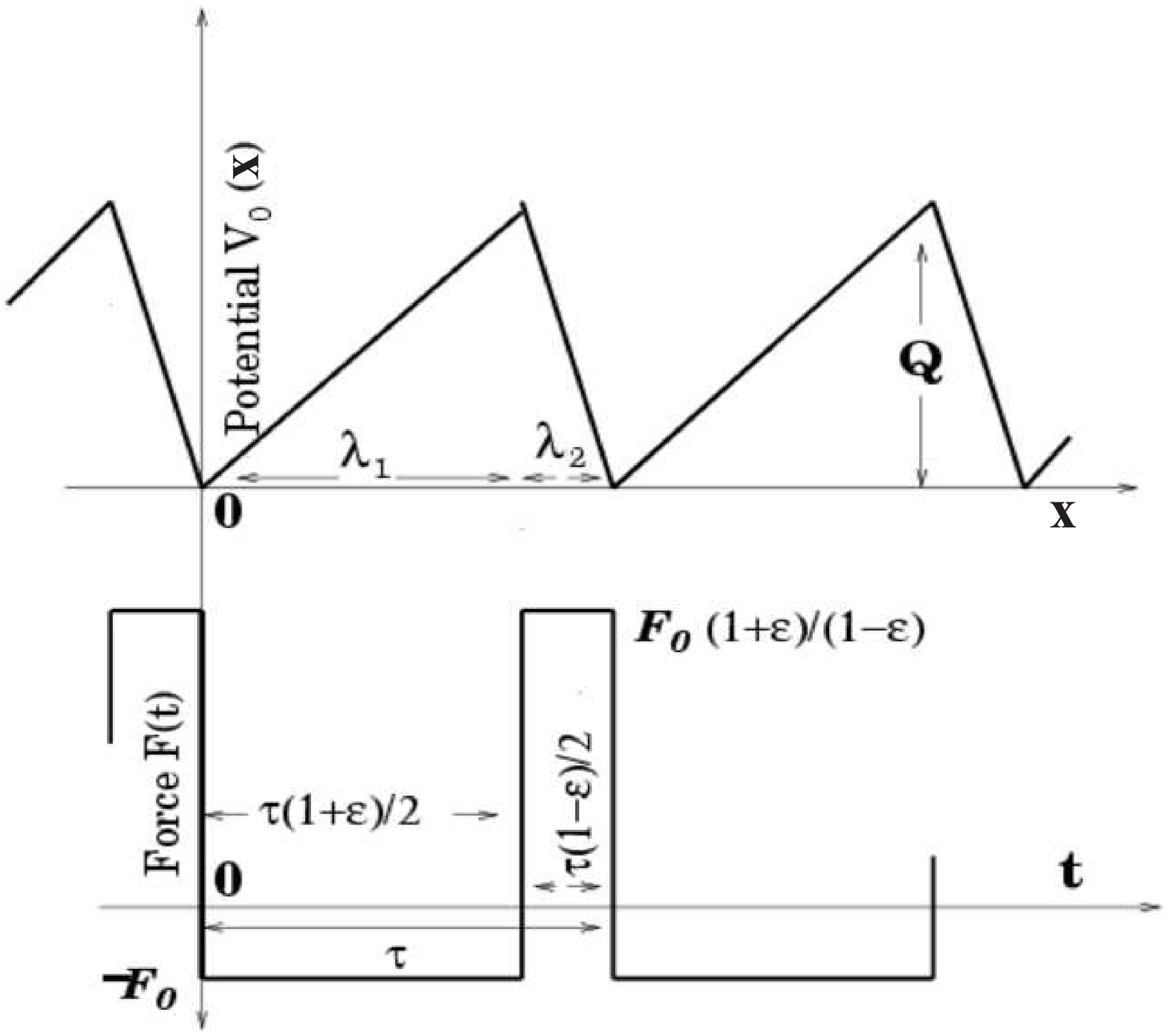}
\caption{Schematic plot of the ratchet potential, $V(x)$ as a function of coordinate $x$ 
and the time asymmetric driving force $F(t)$ as a function of $t$. }
%\label{sawtooth}
\end{center}
\end{figure}

\section{The Model}
We consider the overdamped dynamics of a Brownian particle in a periodic ratchet potential $V(x)$ system in the presence of an external temporal asymmetric driving force $F(t)$. The governed Langevin equation for the overdamped motion of the particle \cite{Dan} is given by
\begin{equation}
 \gamma {\dot{x}} = -V^\prime(x)+F(t)+\xi(t),\label{lang}
 \end{equation}
where $x(t)$ is instantaneous position of the particle and $\gamma$ being the friction coefficient. $\xi(t)$ is the randomly fluctuating Gaussian thermal noise, which satisfies the properties, 
$<\xi(t)\xi(t^\prime)>\,=\,(2\,{k_BT}/\gamma) \delta(t-t^\prime)$ and $<\xi(t)>$=0. The angular bracket $<...>$ denote the ensemble average over all the realizations of noise. 
A schematic diagram of the ratchet potential $V(x)$ and the external drive, $F(t)$ are shown in Fig.1.
\begin{eqnarray}
V(x) &=& \frac{Q}{\lambda_1} x, \,\,\,\,\,\,\,\,x\leq \lambda_1 \nonumber \\
 &=& \frac{Q}{\lambda_2} (1-x),  \,\,\lambda_1<x\leq \lambda,
\end{eqnarray}
where $Q$ is the height of the ratchet potential. $\lambda=\lambda_{1}+\lambda_{2}$, is considered to be the spatial periodicity of the ratchet potential. In the rescaled units, this parameter $\lambda$ is always set to unity. $\Delta=\lambda_{1}-\lambda_{2}$ is the spatial asymmetric parameter which characterizes the spatial asymmetry of the ratchet potential.\\
  
The temporal asymmetric driving force $F(t)$ is given by
\begin{eqnarray}
F(t)&=& \frac{1+\epsilon}{1-\epsilon}\, F,\,\, (n\tau 
\leq t < n\tau+ \frac{1}{2} \tau (1-\epsilon)), \\ \nonumber
    &=& -F,\,\, (n\tau+\frac{1}{2} \tau(1-\epsilon) < t \leq
(n+1)\tau).\label{ft}
\end{eqnarray}
Here $\epsilon$ represents the temporal asymmetric parameter in the driving force and its value ranges from $0$ to $1$. $\tau$($=\frac{2\pi}{\omega})$ denotes the time period of the external drive with $\omega$ being the frequency of the drive. $n=0,1,2....$ is an integer.

This model has been extensively studied both in the adiabatic as well as in the nonidiabatic limit \cite{Raishma, Sahoo, Sahoo1}. The nature of the averaged current, different form of efficiencies of transduction are studied in detail there. In the adiabatic limit and at temperature, $T \rightarrow 0$ limit, the analytical expression for the time averaged current, $<j>$ is obtained in \cite{Raishma}, which is given by
\begin{equation}
<j>=j_{1}+j_{2},
\end{equation}
where $j_{1}$ and $j_{2}$ are the fraction of currents in the positive and negative direction over a time period of $\tau(1-\epsilon)/2$ and $\tau(1+\epsilon)/2$ respectively. $j_{1}$ is the current when the external driving force, $F(t)=(\frac{1+\epsilon}{1-\epsilon})F$ and $j_{2}$ is the current when the external driving force, $F(t)=-F$. 

In the present work, we are mainly interested in the noise induced directed motion and the accompanying diffusive spread or diffusion. The diffusion coefficient, describing the fluctuations across the mean position of the particles, is defined as
\begin{eqnarray}
D=\lim_{t \rightarrow \infty} \frac{1}{2t} [<x^{2}> - <x>^{2}],
\end{eqnarray}

where the angular brackets $<....>$ denote an average over all the realizations of thermal noise. If the average velocity in the asymptotic limit is large and the spread of trajectories is small, the diffusion coefficient is small and the transport is more effective. In our model, the averaged current is same as that of average velocity and is given by
\begin{equation}
<j>=<\frac{x_t-x_0}{t-t_0}>,
\end{equation} 
where $x_t$ and $x_0$ represent the instantaneous position of the particles at time $t$ and $t_{0}$ respectively. To quantify the effectiveness of transport, we introduce the dimensionless Peclet number, 'Pe' which can be defined as

\begin{eqnarray}
Pe=\frac{<j>\lambda}{D},
\end{eqnarray}
where $\lambda$ is the length of the spatial period of the ratchet, which is taken to be unity in our model. 

\begin{figure}[hbp!]
\begin{center}
\input{epsf}
\includegraphics [width=3in,height=2.5in]{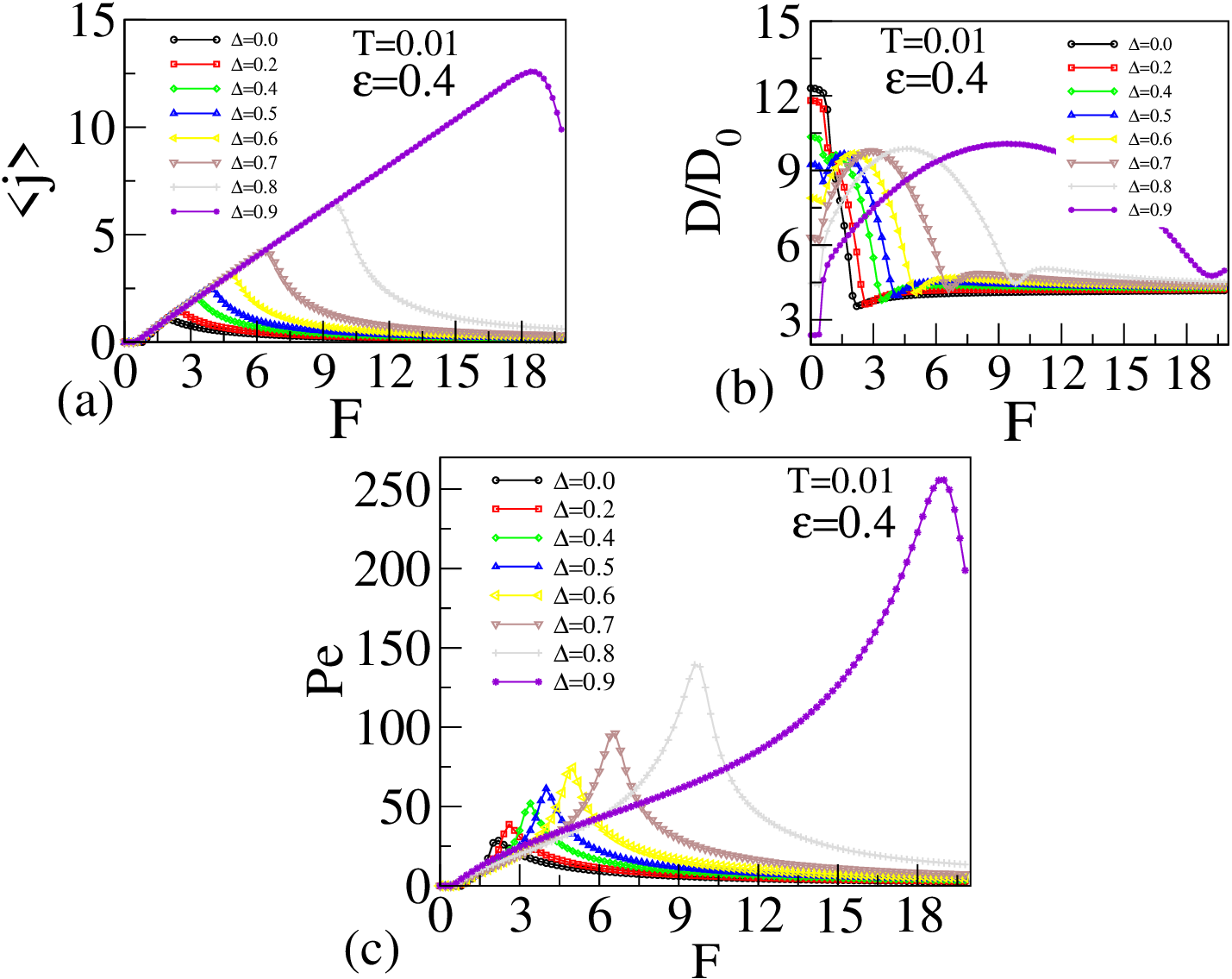}
\caption{Average current, $<j>$ (a), diffusion coefficient, $D/D_{0}$ (b) and the Peclet number, $Pe$ as a function of amplitude of driving, $F$ in the deterministic limit ($T=0.01$) for various values of the spatial asymmetry parameter, $\Delta$. The other fixed parameters are $\epsilon=0.4$ and $\omega0.5$.} 
%\label{sawtooth}
\end{center}
\end{figure}

\section{Numerical Simulation details}
We have used the Huens method algorithim for solving the overdamped Langevin dynamics of our model. From the solution of the dynamics, we use to evaluate the averaged current, $<j>$, diffusion coefficient, $D$ and the Peclet number, $Pe$. We run the simulation upto $10^6 \tau$,  where $\tau$ is one period of the drive. The steady state averages are taken just after ignoring the initial transients upto $10^5 \tau$. The time interval between each time steps is considered to be $0.001$. 

In order to follow all the simulation results, we rescaled all the lengths with respect to the spatial periodicity of the potential, $\lambda$, and energies with respect to the height of the ratchet potential, $Q$. All the physical quantities are in the dimensionless form and the friction coefficient, $\gamma$ is set to be unity. 

\section{Result and Discussion}
We have studied the nature of average current, $<j>$, diffusion coefficient, $D$ and Peclet number, $Pe$ as a function of various physical parameters of the model. This model has been studied in detail both in adiabatic and nonadiabatic limit \cite {Sahoo, Sahoo1}. However, the main interest of those studies were on the efficient operation of the ratchet, hence the physical quantities of interest were different forms of efficiency of energy transduction. In this study we are mainly interested in the coherence of transport in the ratchet, therefore mainly on the diffusion coefficient, $D$ and the Peclet number, $Pe$. In the deterministic limit of the model, i.e., at temperature, $T$ $\rightarrow$ $0$ limit, there exists potential barriers in both forward as well as backward directions. The particle is basically in the locked state and randomly fluctuates across any one single minima of the ratchet for a finite temperature $T>0$, resulting the average velocity to be zero so as the net current. In this limit, the net current can arise from the sum of the contributions from the fraction of current due to the forward forcing of the drive and the fraction of current due to the backward forcing. In the similar way we can evaluate the net diffusion coefficient as the sum of the contributions of diffusion from the forward direction (when the forcing field is applied in the forward direction) and from the backward directions (when the forcing field is applied is in the backward direction) respectively. The forward barriers to the motion of particle in the ratchet disappear only when the forcing amplitude in the forward direction is larger than the force exerted by the forward potential barrier, i.e., when $(\frac{1+\epsilon}{1-\epsilon})>\frac{Q}{\lambda_{1}}$. Similarly, the barriers in the backward direction disappear only when the backward forcing amplitude is larger than the force exerted by the potential barrier in the backward direction, i.e., when $F>\frac{Q}{\lambda_{2}}$. Hence, two crictical forces, $F_{c1}=\frac{Q}{\lambda_{1}}\frac{(1+\epsilon)}{(1-\epsilon)}$ and $F_{c2}=\frac{Q}{\lambda_{2}}$, exist in the ratchet potential beyond which there are no barriers to motion of particle in both forward as well as backward direction. As a result there exists a finite positive current in the ratchet, only when the forcing amplitude is in between the crictical fields, $F_{c1}$ and $F{c2}$.

\begin{figure}
%[hbp!]
\begin{center}
\includegraphics [width=3in,height=2.5in]{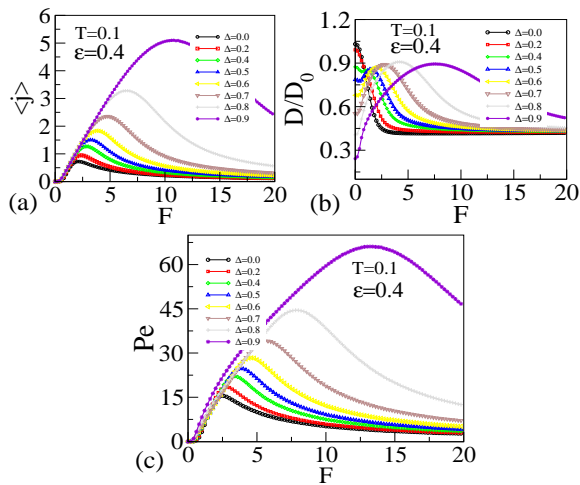}
\caption{Average current, $<j>$ (a), diffusion coefficient, $D/D_{0}$ (b) and the Peclet number, $Pe$ as a function of amplitude of driving, $F$ for various values of the spatial asymmetry parameter, $\Delta$. The other fixed parameters are $T=0.1$, $\epsilon=0.4$ and $\omega=0.5$.}
%\label{sawtooth}
\end{center}
\end{figure}

In fig.2, we have plotted the average current,$<j>$, diffusion coefficient, $D/D_{0}$ and the Peclet number, $Pe$ as a function of amplitude of driving, $F$ for various values of the spatial asymmetry parameter, $\Delta$ in the deterministic limit. $D_{0}=k_{B}T/\gamma$ refers to the free diffusion coefficient. We fix the other parameters of the model same as that where the ratchet shows an efficient performance (resulting large values of both Stokes efficiency and thermodynamic efficiency). For both symmetric as well as asymmetric potentials, both the average current, $<j>$ (fig.1(a)) and $<Pe>$ (fig.1(c)) show peaking behaviors as a function of amplitude of driving, $F$. Interestingly the Peclet number, $Pe$ follows the same nature of the $<j>$ like the Stokes efficiency and shows a maximum at the same value of forcing amplitude, $F$, at which the average current, $<j>$ also show a maximum. For forcing amplitude, $F<F_{c_{1}}$, there are potential barriers to the motion of particle in both the forward as well as backward direction and results a zero net current in the ratchet. In this limit, the particle gets trapped in one of single well of the ratchet potential and randomly fluctuates across any one minima of the well and results a finite diffusion in the presence of a bias in either of the forward or backward direction. Therefore the effective diffusion coefficient ($D/D_{0}$) (fig.1(b)) shows a finite positive value even at the zero value of amplitude of driving. This is due to the presence of a bias, i.e., finite temperature ($T=0.01$) in the ratchet. With further increase in amplitude of driving, the forward barriers disappear at $F=F_{c_{1}}$ and the particle progresses more towards the forward direction without any back turn. As a result the average current, $<j>$ and $Pe$ increases with $F$. However, at the same time the effective diffusion coefficient, $D/D_{0}$ decreases with $F$. When the forcing amplitude, $F$ becomes $F_{c_{2}}=\frac{Q}{\lambda_{2}}$, the potential barriers in the backward direction disappear and the particle starts moving in the backward direction, resulting a negative contribution towards the average current, $<j>$. However, the diffusion in the backward direction increases. Both the average current, $<j>$ and the peclet number shows a maximum at $F=F_{c_{2}}$ where as the diffusion coefficient $D/D_{0}$ passes through a minimum in the range of $F$ around $F=\frac{Q}{\lambda_{2}}$, exactly where the Peclet number shows a maximum, supporting a strong enhancement of transport coherence in this parameter regime. With increase in the spatial asymmetry ($\Delta$) in the ratchet potential, both $<j>$ and $Pe$ increases and the maximum values shift towards the larger values of $F$. At the same time the minimum values in the diffusion curve also shift towards right and shows a larger  spreading for the values of $F$ in between $F_{c_{1}}$ and $F_{c_{2}}$. For both symmetric as well as asymmetric ratchet potentials, the diffusion coefficient $D/D_{0}$ saturates for larger $F$ values and is no more affected by the amplitude of driving. This is because, for larger $F$ values, the barriers in both the directions disappear, as a result the particle behaves like a free particle in the absence of any potential barrier. This is the case of a one dimensional free diffusion and $D \sim D_{0}$. However, both the average current, $<j>$ and Peclet number, $Pe$ approaches zero values in this limit. From fig.1(c), we noticed that the Peclet number, $Pe$ is very high (around $270$) for a larger spatial asymmetry parameter ($\Delta=0.9$) in the ratchet potential. We conclude that in this particular ratchet model, in the adiabatic regime of operation, i.e., for low frequency of driving, the larger spatial asymmetry in the potential helps in enhancing the coherence of transport in the deterministic limit.

In fig.3, the nature of average current, $<j>$ (fig.3(a)), diffusion coefficient, $D/D_{0}$ (fig.3(b)) and the Peclet number, $Pe$ (fig.3(c)) as a function of amplitude of driving, $F$ are shown for various values of the spatial asymmetry parameter, $\Delta$ in the ratchet potential and for a higher temperature, $T=0.1$. They show the similar behavior as a function of amplitude of driving as already seen in fig.2. However, the peaks in both $<j>$ and $Pe$ as a function of $F$ as well as the minimum in $D/D_{0}$ curves get suppressed. At the same time the peaks in $<j>$ and $Pe$ get broader. This is because, larger temperature facilitates the escape of particle in the backward direction as well even in the presence of potential barriers. The particle spreads over both the forward as well as backward directions for a larger temperature resulting a decrease in the average current, $<j>$ and the Peclet number, $Pe$. 

\begin{figure}
%[hbp!]
\begin{center}
\includegraphics [width=3in,height=2.5in]{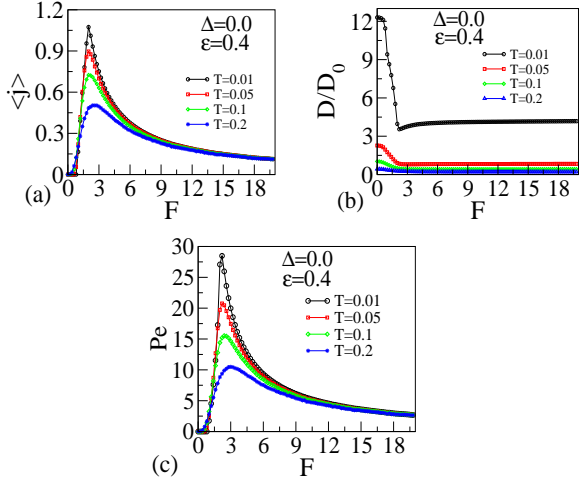}
\caption{The average current, $<j>$, the diffusion coefficient, $D/D_{0}$ and the Peclet number, $Pe$ as a function of amplitude of driving, $F$ in (a), (b) and (c) respectively for a symmetric potential and for different values of temperature, $T$,. The other fixed parameters are $\Delta=0.0$, $\epsilon=0.4$ and the frequency of driving, $\omega=0.5$.}
%\label{sawtooth}
\end{center}
\end{figure}
Next in fig.4, we have studied the average current, $<j>$, the diffusion coefficient, $D/D_{0}$, and the Peclet number, $Pe$ as a function of amplitude of driving, $F$ for a symmetric potential and for various values of temperature, $T$. We observe that, both $<j>$ and $Pe$ as a function of $F$ decrease  with increase in temperature and the peak values both in $<j>$ (fig.4(a)) and $Pe$ (fig.4(c)) get suppressed and the peaks get broader. Also the value of finite diffusion at zero amplitude of driving (fig.4(b)) gets reduced. The crictical value of the ampitude of drive at which both the current and Peclet number show a maximum, shift towards larger $F$ values with increase in temperature. Similarly the minimum in the diffusion coefficient, $D/D_{0}$ as a function of $F$ curve, also shift towards right. This is due to the reason that, with increase in temperature, the thermal noise facilitates the progress of particle even in the backward direction, giving a negative contribution towards the average current, effective diffusion and in the Peclet number. As a result effectively larger forces are required to overcome the barriers to the motion of the particle in the ratchet. In this parameter regime of ratchet operation, the higher temperature supresses the quality of transport and does not favor in the coherent performance of the ratchet.

\begin{figure}
%[hbp!]
\begin{center}
\includegraphics [width=3in,height=2.5in]{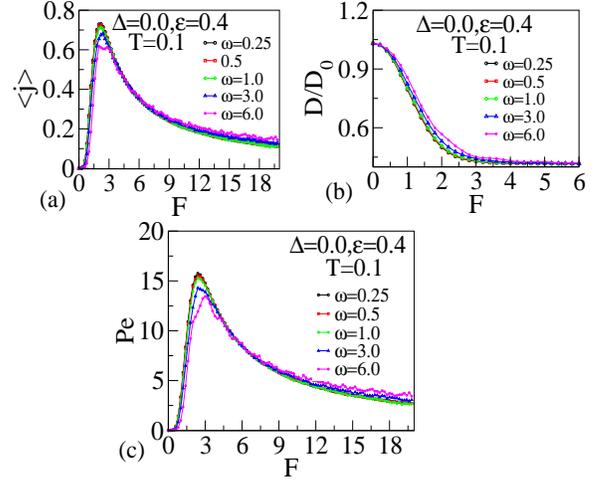}
\caption{The average current, $<j>$, the diffusion coefficient, $D/D_{0}$ and the Peclet number, $Pe$ as a function of amplitude of driving, $F$ in (a), (b) and (c) respectively for a symmetric potential and for different values of frequency of driving, $\omega$. The other fixed parameters are $\Delta=0.0$, $\epsilon=0.4$ and the temperature $T=0.1$.}
%\label{sawtooth}
\end{center}
\end{figure}

The nature of the average current, $<j>$, Peclet number, $Pe$ and the difussion coefficient, $D/D_{0}$  as a function of amplitude of drive, $F$ is studied in fig.5, for a symmetric potential and for various values of frequency of drive, $\omega$. For a particular frequency of drive, the average current, $<j>$ (fig.5(a)) and the Peclet number, $Pe$ (fig.5(c)) show a peaking behavior as a function of amplitude of drive. However, the Peclet number, $Pe$ follows the same nature of average current, $<j>$ for all values of frequency of drive. With increase in the frequency of drive, the peak values both in $<j>$ as well as in $<Pe>$ curve decrease and at the same time the peaks shift towards larger $F$ values. This is because with increase in the frequency of drive, the forcing amplitude varies so fast that the particle is not able to take the advantage of force in either of the forward or backward directions and the motion of the particle towards the nearest neighbouring minima of the ratchet potential gets suppressed. This results a decrease in the average current as well as in the Peclet number. The increase of frequency of driving degrades the coherence of transport in the ratchet. For a particular frequency of driving, the diffusion coefficient, $D/D_{0}$ as a function of $F$ (fig.5(b)) shows a decreasing behavior with $F$. Also the difussion coefficient, $D/D_{0}$ is very slightly affected or not at all affected by modulating the frequency drive. This is because for high frequency of drive, the forcing amplitude varies so fast that the particle doesn't feel the influence the forcing. The particle as if stays at one position only without progressing in either of the directions of the ratchet as a result the diffusion coefficient is not much affected by modulating the frequency of drive.

\begin{figure}
%[hbp!]
\begin{center}
\includegraphics [width=3in,height=2.5in]{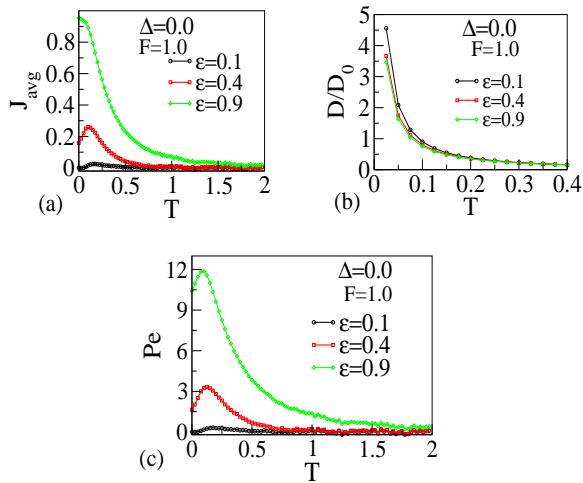}
\caption{The average current, $<j>$, the diffusion coefficient, $D/D_{0}$ and the Peclet number, $Pe$ as a function of temperature, $T$ in (a), (b) and (c) respectively for a symmetric potential and for different values of temporal asymmetry parameter, $\epsilon$,. The other fixed parameters are $\Delta=0.0$, $\omega=0.5$ and $F=1.0$.}
%\label{sawtooth}
\end{center}
\end{figure}

\begin{figure}
%[hbp!]
\begin{center}
\includegraphics [width=3in,height=2.5in]{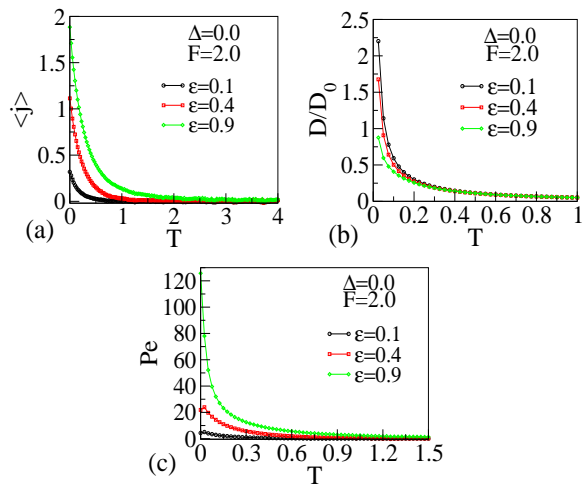}
\caption{The average current, $<j>$, the diffusion coefficient, $D/D_{0}$ and the Peclet number, $Pe$ as a function of temperature, $T$ in (a), (b) and (c) respectively for a symmetric potential and for different values of temporal asymmetry parameter, $\epsilon$,. The other fixed parameters are $\Delta=0.0$, $\omega=0.5$ and $F=2.0$ respectively.}
%\label{sawtooth}
\end{center}
\end{figure}

In fig.6 and fig.7, we have plotted the average current, $<j>$, diffusion coefficient, $D/D_{0}$ and the peclet number, $Pe$ as a function of temperature in (a), (b) and (c) respectively for a symmetric potential and for various values of the temporal asymmetry parameter, $\epsilon$ (for $F=1.0$ in fig.6 and for $F=2.0$ in fig.7). Here we observe that the Peclet number, $Pe$ shows a peaking behaviour as a function of temperature unlike the case of previously studied models \cite{Roy}. It follows the same nature of average current like the Stokes efficiency. Both the average current, $<j>$ and the Peclet number, $Pe$ show a finite positive value even for the values of temperature, $T$ in the $T \rightarrow 0$ limit. This is due to the disappearance of forward potential barrier in the presence of a temporal asymmetry in the driving. These values decrease with further increase in the temporal asymmetry parameter, $\epsilon$. The Peclet number decreases for high temperature implying that the transport coherence decreases with increase in noise strength. For low values of noise strength, i.e.,in the $T \rightarrow 0$ limit, a high Peclet number around $Pe=270$ is observed (fig.2(c)) in this model. However, with further increase in the value of noise strength, the escape of particle becomes faster yielding a slightly more regular process than a simple random walk and for high temperatures, the backward transition occurs and the ratcheting effect is weakened. As a result the Peclet number decreases. Furthermore, the difussion coefficient, $D/D_{0}$ (fig. 6(b)) as a function of temperature shows a decreasing behaviour. Interestingly it shows a $\frac{1}{T}$ like behaviour, where $T$ is the temperature. This reflects the case of a one dimensional random walk. In this parameter regime, the particle behaves like a one dimensional random walker, where the fluctuations in it's position $x$, $\sigma_{x}=<x^{2}>$ is proportional to the square root of the total time taken by it.

\begin{figure}
%[hbp!]
\begin{center}
\includegraphics [width=3in,height=2.5in]{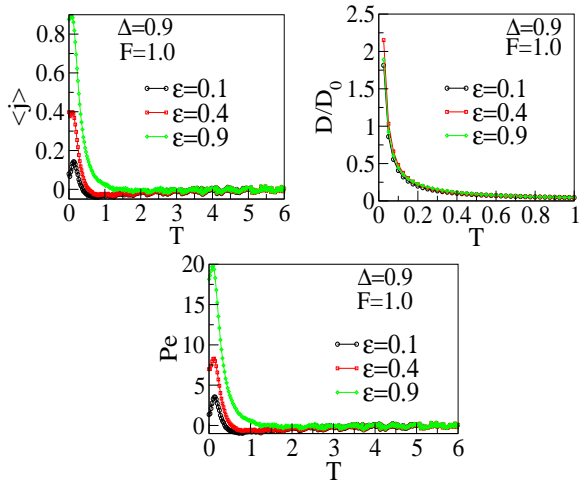}
\caption{The average current, $<j>$, the diffusion coefficient, $D/D_{0}$ and the Peclet number, $Pe$ as a function of temperature, $T$ in (a), (b) and (c) respectively for an asymmetry potential ($\Delta=0.9$) and for different values of temporal asymmetry parameter, $\epsilon$,. The other fixed parameters are $\omega=0.5$ and $F=1.0$.}
%\label{sawtooth}
\end{center}
\end{figure}

In fig.8, the nature of average current, $<j>$, the diffusion coefficient, $D/D_{0}$ and the Peclet number, $Pe$ as a function of temperature, $T$ are studied for an asymmetry potential and for various values of the temporal asymmetry parameter, $\epsilon$. Similar behaviors of $<j>$, $D/D_{0}$, and $Pe$ as a function of temperature, $T$ have been observed as already seen in fig.6. However, both the average current, $<j>$ (fig.8(a)) and Peclet number, $Pe$ (fig.8(c)) show larger values as compared to the case of a symmetric ratchet potential. The diffusion coefficient, $D/D_{0}$ curve as a function temperature (fig.8(b)) shows a more stiffer behavior than the case of symmetric potential.

\section{Conclusion}
We have studied the overdamped Langevin dynamics of a Brownian particle in a temporal asymmetric rocked ratchet model. This model is already studied in detail in both adiabatic and nonadiabatic limits for transport current and different efficiencies \cite{Sahoo, Sahoo1}. It was observed that in the adiabatic regime of operation, this particular ratchet model shows a high efficient performance in a suitable range of parameter space of the model \cite{Sahoo}. However, the quality or reliability of transport was not reported. In this study we took interest in the coherence or reliability of transport for the same parameter regime. We observe that in the deterministic limit and in the adiabatic regime of ratchet operation, the transport coherence is very high for a larger spatial asymmetry in the ratchet potential. The transport coherence decreases with increase in frequency of driving as well as for high temperatures. From the behaviour of diffusion coefficient as a function of temperature, it is reflected that the motion of the particle in the ratchet is like a one dimensional random walker. The ratcheting behavior is more pronounced for a moderate temperature and gets weakened for high temperatures. 

\section{Acknowledgment}
MS acknowledges the funding from the INSPIRE faculty award (IFA-13 PH-66) by the Department of Science and Technology, Govt. of India. AMJ thanks DST, India for financial support (via J C Bose fellowship).

%\section{References}

\end{document}